\documentclass[12pt]{article}
\usepackage[utf8]{inputenc}
\usepackage{heck}
\usepackage{amsmath,amssymb}
\usepackage{color}
\definecolor{myblue}{rgb}{.97,.97,1}
\usepackage{empheq}
\usepackage{CJK}
\usepackage{hyperref}
\usepackage{url}
\usepackage{graphicx}  

\graphicspath{{./Figures_orig/}}

\newlength\mytemplen
\newsavebox\mytempbox

\makeatletter
\newcommand\mybluebox{%
    \@ifnextchar[
       {\@mybluebox}%
       {\@mybluebox[0pt]}}

\def\@mybluebox[#1]{%
    \@ifnextchar[
       {\@@mybluebox[#1]}%
       {\@@mybluebox[#1][0pt]}}

\def\@@mybluebox[#1][#2]#3{
    \sbox\mytempbox{#3}%
    \mytemplen\ht\mytempbox
    \advance\mytemplen #1\relax
    \ht\mytempbox\mytemplen
    \mytemplen\dp\mytempbox
    \advance\mytemplen #2\relax
    \dp\mytempbox\mytemplen
    \colorbox{myblue}{\hspace{1em}\usebox{\mytempbox}\hspace{1em}}}
\makeatother  

\newcommand{\sfy}{\ensuremath{\mathsf{y}}}

\newcommand{\Teichmuller}{Teichm\"{u}ller }
\newcommand{\Schrodinger}{Schr\"{o}dinger }

\begin{document}
\begin{CJK*}{UTF8}{min}

\title{Quantum Trilogy:\\ Discrete Toda, Y-System and Chaos}

\authors{\centerline{Masahito Yamazaki (山崎雅人)}}

\institution{IPMU}{\centerline{Kavli IPMU (WPI), University of Tokyo, Kashiwa, Chiba 277-8583, Japan}}
\institution{Harvard}{\centerline{Center for the Fundamental Laws of Nature,
 Harvard University, Cambridge, MA 02138, USA}}

\abstract{
We discuss a discretization of the quantum Toda field theory associated with a semisimple finite-dimensional Lie algebra or a tamely-laced infinite-dimensional Kac-Moody algebra $G$, generalizing the previous construction of discrete quantum Liouville theory for the case $G=A_1$. The model is defined on a discrete two-dimensional lattice, whose spatial direction is of length $L$. In addition we also find a ``discretized extra dimension'' whose width is given by the rank $r$ of $G$, which decompactifies in the large $r$ limit. For the case of $G=A_N$ or $A_{N-1}^{(1)}$, we find a symmetry exchanging $L$ and $N$ under appropriate spatial boundary conditions. The dynamical time evolution rule of the model is a quantizations of the so-called Y-system, and the theory can be well-described by the quantum cluster algebra. We discuss possible implications for recent discussions of quantum chaos, and comment on the relation with the quantum higher \Teichmuller theory of type $A_N$.
}


\maketitle 
\end{CJK*}
\tableofcontents

\section{Introduction}

In celebrated papers in 1967 \cite{Toda:1967JPSJ,Toda:1967JPSJ_2}, Prof.\ Morikazu Toda introduced the now-famous Toda lattice. Since then, the Toda lattice became one of the most important and well-studied models of integrable models.  The author of the present paper himself has long been fascinated by the subject, ever since his first encountered the model in an excellent book by Toda \cite{Toda_Book_NL}.

In this paper, we discuss a discretization of the two-dimensional quantum Toda field theory associated with a 
symmetry algebra $G$, which is either a finite-dimensional semisimple Lie algebra or an
infinite-dimensional tamely-laced Kac-Moody algebra (the tamely-laced condition will be explained later in \eqref{tame_laced}).
We consider the $(1+1)$-dimensional spacetime, and 
discretize both the time and spatial directions. The model is thus defined on a two-dimensional lattice.  When $G=A_1$, this reduces to the discrete quantum Liouville theory, discussed in \cite{Faddeev:1985gy,Faddeev:1993pe,Faddeev:1997hp,Faddeev:2000if,Faddeev:2002ms}.

There are several motivations for the quantum discrete Toda theory.

First, such a discretization is suitable for analyzing the Toda field theory as an integrable model.
While Toda (and especially Liouville) theory has been discussed extensively in the literature,
most papers resort to conformal field theory (CFT) techniques. However, Toda theory is also integrable,
where the integrable model techniques \cite{Faddeev:1987ph,Faddeev:1996iy} in discrete spin chains models should apply. Discrete model also serves as a UV regularization of the continuum theory.

Second, discrete Toda theory has been the birthplace of the (non-compact) quantum dilogarithm function \cite{Faddeev:1993rs}, which later appeared in a number of different contexts in physics and mathematics,
including quantum \Teichmuller theory \cite{Chekhov:1999tn,KashaevQuantization,FockGoncharovHigher}, 
complex Chern-Simons theory \cite{Hikami:2006cv,Dimofte:2009yn,Andersen:2011bt} and finally in 3d $\mathcal{N}=2$ supersymmetric theory \cite{Hama:2011ea}, as related by the 3d/3d correspondence \cite{Terashima:2011qi,Dimofte:2011ju} (see \cite{Lee:2013ida,Cordova:2013cea} for derivation, also \cite{Yagi:2013fda})
and the gauge/YBE correspondence \cite{Terashima:2012cx,Yamazaki:2012cp,Yamazaki:2013nra} (see also \cite{Bazhanov:2007mh,Spiridonov:2010em}).

More recently, discrete Liouville theory has been studied in the context of quantum chaos \cite{Turiaci:2016cvo}.
Namely, it is a concrete $(1+1)$-dimensional lattice model saturating the 
conjectured bound \cite{Maldacena:2015waa} for chaos, and hence can be thought of as a higher-dimensional counterpart of the $(0+1)$-dimensional model proposed by Sachdev, Ye, and Kitaev \cite{Sachdev:1992fk,Kitaev}. Discrete Toda theory is a natural generalization
whose semiclassical holographic dual (in the large central charge) contains particles with spin greater than $2$,
which is of interest in view of the recent constraints on such theories from causality \cite{Camanho:2014apa} and quantum chaos \cite{Maldacena:2015waa,Perlmutter:2016pkf}.

Readers should keep in mind that in the literature there has been many papers
on discretizations of classical Toda equations, as early as in the seventies \cite{Ablowitz,Hirota1}.
The connection of the discrete classical Toda system with the classical Y-system (see \cite{Krichever:1996qd,Kuniba:1995gi}) and cluster algebra, to be discussed below, is also known in the literature. 
Furthermore {\it quantum}  aspects of discrete Liouville theory  was discussed in \cite{Faddeev:1985gy,Faddeev:1993pe,Faddeev:1997hp,Faddeev:2000if,Faddeev:2002ms}, and discrete $A_N$ Toda theory in \cite{Kashaev:1995mq}\footnote{The latter reference also discussed the $N\to \infty$ limit, to be discussed in this paper.}.\footnote{The references \cite{Bytsko:2009mg,Meneghelli:2015sra}
also discuss discretization of the quantum Liouville and Toda theory.}

In this paper, we reformulate many of these pioneering results in the literature
in modern machinery of quantum cluster algebras. Formulated this way, much of the involved computations
\cite{Faddeev:1985gy,Faddeev:1993pe,Faddeev:1997hp,Faddeev:2000if,Faddeev:2002ms,Kashaev:1995mq}
are replaced by rather simple combinatorics of quivers and mutations. We discuss Toda theories associated with more general symmetry groups $G$ and with more general boundary conditions than known in the literature, discuss subtle relations with quantum higher \Teichmuller theory of \cite{FockGoncharovHigher}, and point out the connection with recent exciting developments of quantum chaos.

In the rest of this paper, in Sec.~\ref{sec.model} we first we 
describe the model and comment on some of its properties.
We then comment on the possible implications to quantum chaos in Sec.~\ref{sec.chaos}
and clarifies a relation connection with \Teichmuller theory in Sec.~\ref{sec.Teichmuller},
We conclude this paper with some closing remarks in Sec.~\ref{sec.conclusion}.

\section{Discrete Toda Theory}\label{sec.model}

In this section we introduce and motivate the discrete Today theory (see Appendix \ref{subsec.continuum} for continuum theory and for some notations).

As we will comment in Sec.~\ref{subsec.non-simply-laced}, the formulation below of the discrete models will work both for a finite-dimensional Lie algebra $G$ as well as an infinite-dimensional tamely-laced Kac-Moody algebra. This will include most of the infinite-dimensional affine Lie algebras.
In order to simplify the presentation, for the most part we choose the symmetry algebra $G$ to be a simply-laced simple finite-dimensional Lie algebra. We will comment on the non-simply-laced case and the infinite-dimensional case in Sec.~\ref{subsec.non-simply-laced}.

\subsection{Discrete Model}

Let us now come to the definition of the statistical mechanical model.
We choose to use the Hamiltonian viewpoint, first quantize the theory 
along a fixed time slice, and then consider its time evolution. 

\subsubsection{Dynamical Variables }\label{subsec.dynamical}

Let us first consider the situation with a fixed time slice (say at time $t$). Here, we have a discrete lattice whose lattice points are labeled by a pair of integers.

First, we have an integer $a=1,\ldots, r$, labeling the ``internal symmetry'' of the Toda theory.
Second, we have another integer $m$ for the spatial directions, 
which runs from $1$ to $L$, where $L$ specifies the length of the spatial direction.
This integer $L$ is taken to be infinity in the continuum limit.

In the following we sometimes combine $(a, m)$ into a single index, which we denote by 
$i, j,\cdots I$, where $I$ is the index set $I=\{1, \ldots, r\} \times \{1, \ldots, L\}$.

For each vertex $(a, m)$ we associate a variable $\sfy_i(t)=\sfy^a_m(t)$; these will be the dynamical variables of the theory. One the boundary of the spatial directions,
we can either consider either
\begin{itemize}
\item
(P): a periodic boundary condition, in which case $m$ is considered to be 
modulo $L$: $\sfy^a_{m+L}(t)=\sfy^a_m(t)$, in which case the spatial direction is a circle, or
\item
(F): a fixed boundary condition, in which case we allow the integer $m$ to take values in $0$ and $L+1$, and fix their values to be $\sfy^a_{m=0}(t)^{-1}=\sfy^a_{m=L+1}(t)^{-1}=0$.
\end{itemize}
In each of these cases we can choose to take $L=\infty$, in which case we have an
infinite chain.

\subsubsection{Commutation Relations}\label{subsec.commutation}

At the fixed time slice at time $t$, let us next set a commutation relation
between the $\sfy_m^a(t)$'s.

This is determined by the symmetry algebra $G$, as well as another 
Lie algebra $G'$, which is defined to be 
\begin{align}
G'=\begin{cases}
A_L & (F) \\
A_{L-1}^{(1)} & (P) 
\end{cases}
\;,
\end{align}
depending on the boundary condition.
We again allow $L=\infty$, leading to $G'=A_{\infty}$ or $A_{\infty}^{(1)}$.

\begin{figure}[htbp]
\centering\includegraphics[scale=0.6]{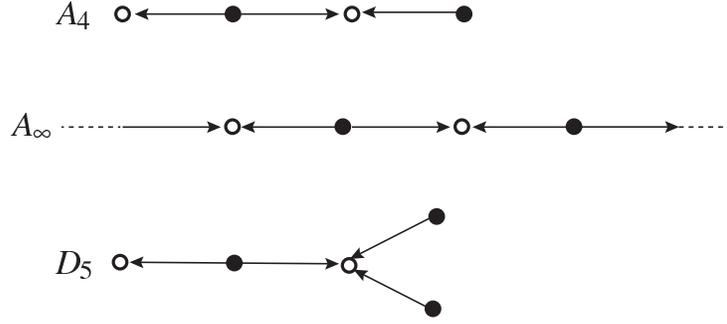}
\caption{Dynkin diagram for $G=A_4, A_{\infty}$ and $D_5$.
The vertices of the diagram can be colored bipartite, and we oriented the arrows 
to be orienting from black to white.
}
\label{fig.dynkin}
\end{figure}

\begin{figure}[htbp]
\centering\includegraphics[scale=0.6]{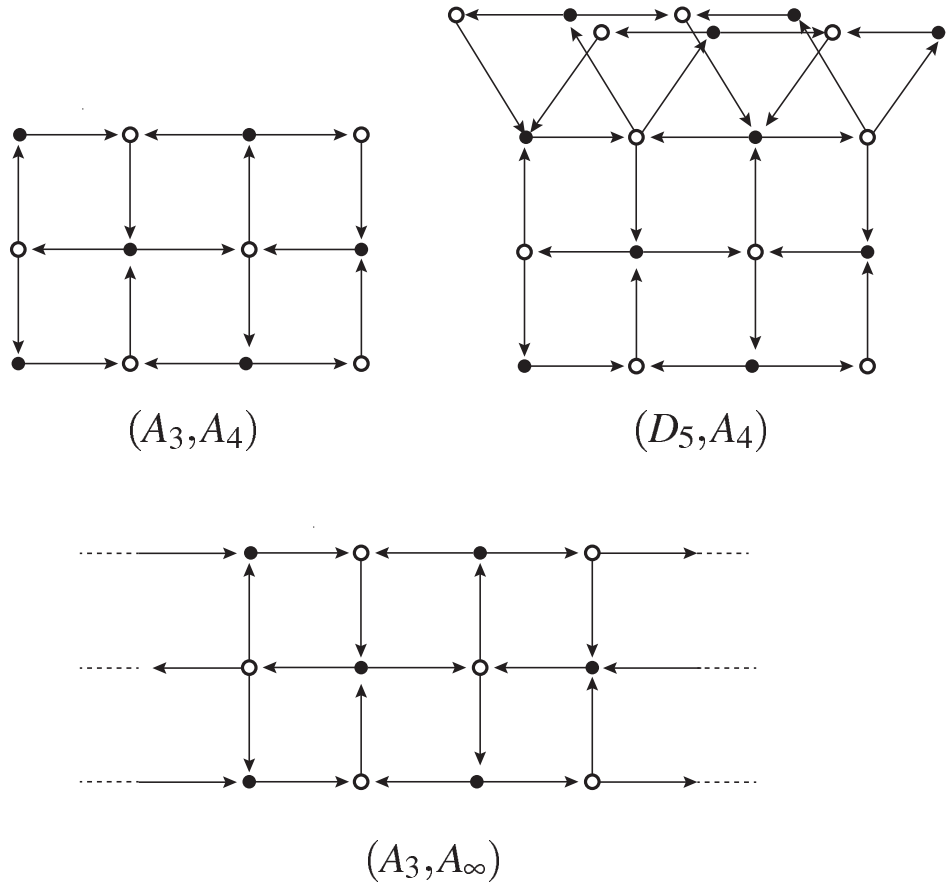}
\caption{The commutation relations for our variables $\sfy^a_m(t)$ are determined from the 
oriented graph (quiver) for the square product $Q=G\, \Box\, G'$. As shown in this figure,
the square product $\Box$ is obtained by combining the Dynkin diagrams for $G$ and that for $G'$.
}
\label{fig.quiver}
\end{figure}

Now that we have a pair of Lie algebras $G$ and $G'$, we can define a 
two-dimensional quiver by combining the Dynkin diagram $Q_G$ for $G$ and 
that ($Q_{G'}$) for $G'$ \cite{KellerPeriodicity}, see Figs.~\ref{fig.dynkin} and \ref{fig.quiver}.
This quiver is called the square product of $Q_G$ and $Q_{G'}$,
and is often denoted by
\begin{align}
Q=Q_{G}\, \Box\, Q_{G'} \;.
\label{QGG}
\end{align}
We also called this the $(G, G')$-quiver.\footnote{Such a quiver has been 
discussed in the context of 4d $\mathcal{N}=2$ \cite{Cecotti:2010fi} and 4d $\mathcal{N}=1$ \cite{Heckman:2012jh}
theories.}

The definition of the square product is hopefully clear from Fig.~\ref{fig.quiver},
but we can give a more formal definition \cite{KellerPeriodicity}. For this, let us first 
write choose a bipartite coloring of the Dynkin diagrams $Q_G$ and $Q_{G'}$,\footnote{For periodic boundary condition, we assume $L$ to be even, so that $Q_{G'}$ is bipartite.}
and we orient their edges so that all the arrows are starting from (ending at)
white (black) vertices.
Then we first form the
tensor product $Q_G\otimes Q_{G'}$ of $G$ and $G'$,
whose the vertices are the pair $(a,m)$ of the vertices $a$ ($m$) for $Q_G$ ($Q_{G'}$),
and the number of arrows from $(a,m)$ to $(b,n)$ is 
\begin{itemize}\parskip=-2pt
\item zero if $a\ne b$ and $m\ne n$
\item equals the number of arrows from $m$ to $n$ if $a=b$
\item equals the number of arrows from $a$ to $b$ if $m=n$
\end{itemize}
Then the square product $G\, \Box\, G'$ is obtained by reversing all arrows in the full subquivers of the form $\{ a \}\times Q_{G'}$ and  $Q_G\times \{ m \}$ where $a$ is a sink of $Q_G$ and $m$ is a source of $Q_{G'}$.

Since a vertex of the Dynkin diagram of $G$ (of $G'$) are labeled by 
an index $a$ ($m$), a vertex of $Q$ is labeled by $i=(a, m)\in I$,
the set labeling the $\sfy_i(t)$'s.

We can now state the commutation relation amongst $\sfy$'s at time $t$:
\begin{empheq}[box={\mybluebox[7pt]}]{equation} 
\sfy_j(t)\, \sfy_i(t)=q^{2\, Q_{ij}} \sfy_i(t) \, \sfy_j(t) \;,
\label{xCCR}
\end{empheq}
where we defined the anti-symmetric matrix $Q_{i,j}$ (with $i,j\in I$) 
by the relation
\begin{align}
Q_{ij}:=\#\{\text{arrows from $i$ to $j$}\}-
\#\{\text{arrows from $j$ to $i$}\} \;.
\label{Q_def}
\end{align}

The non-commutativity parameter $q=e^{i\pi b^2}$ in \eqref{xCCR} is the quantum parameter, namely $\pi b^2$ plays the role of the Planck constant. 
This parameter $b$ is to be identified with the 
same parameter $b$ appearing in Sec.~\ref{subsec.continuum}
for the continuum theory. 
Let us here remark that the semiclassical limit $c\to \infty$ in the continuum theory is 
$b\to 0$, namely $q\to 1$, which is also the semiclassical limit of the 
discrete model.\footnote{The Toda theory is known to have a symmetry $b\to b^{-1}$ (recall that $G$ is simply-laced in this subsection),
as is manifest in the formula \eqref{c_Toda}. This means that we should consider the so-called modular double, and should consider two copies of \eqref{xCCR}, one with $b$ and another with $b^{-1}$ (see {\it e.g.}\,\cite{Gang:2015wya} for some more details). While this is important for some considerations of the Toda theory, 
this is not necessary for the semi-classical consideration of the theory $b\to 0$, including the 
application to quantum chaos discussed in Sec.~\ref{sec.chaos}.}

Note that in logarithmic variables ($\sfy_i=\exp(\mathsf{Y}_i)$), 
the commutation relation \eqref{xCCR} can be written as
\begin{align}
[\mathsf{Y}_i, \mathsf{Y}_j] = i \pi b^2 \, Q_{ji} \;,
\label{YYQ}
\end{align}
After some linear change of basis this reduces to the canonical commutation relations,
which we can easily quantize following the standard canonical quantization procedure.

Note that the commutation relation \eqref{xCCR} allows for a finite-dimensional cyclic representation when $q^2$ is a root of unity; we then have a quantum mechanical model where everything is regularized to be finite. This can be still consistent with the semiclassical limit
when we take $q=\exp(i\pi /M)$ with $M$ a large integer \cite{Turiaci:2016cvo}.

\subsubsection{Time Evolution}\label{subsec.evolution}

Let us now consider the time evolution. This is described by the following equation, 
which we call the quantum Y-system (for simply-laced $G$):
\begin{empheq}[box={\mybluebox[7pt]}]{align} 
\sfy_m^a(t+1)\, \sfy_m^a(t-1) 
&=
\frac{
\prod_{b\ne a} (1+\sfy_m^{b}(t))^{-C_{ab}}
}{
\prod_{n\ne m}(1+\sfy_{n}^a(t)^{-1})^{-C'_{mn}} 
}\nonumber \\
&= 
\frac{
\prod_{b: C_{ab}=-1} (1+q\, \sfy_m^b(t))
}{
(1+q \,\sfy_{m-1}^a(t)^{-1}) (1+q\, \sfy_{m+1}^a(t)^{-1})
}\;,
\label{eq.evolve}
\end{empheq}
where $C_{ab}$  ($C'_{mn}$) is the Cartan matrix\footnote{For example, for $G=A_N$ we have 
\begin{align}
C_{ab}=
\begin{cases}
2 & (a=b) \\
-1 & (|a-b|= 1) \\
0 & (\textrm{otherwise})
\end{cases}\;.
\end{align}
} for $G$ (for $G'$)
 and we have defined $\sfy_m^{0}(t)=\sfy_m^{r+1}(t)=0$. 

\begin{figure}[htbp]
\centering\includegraphics[scale=0.6]{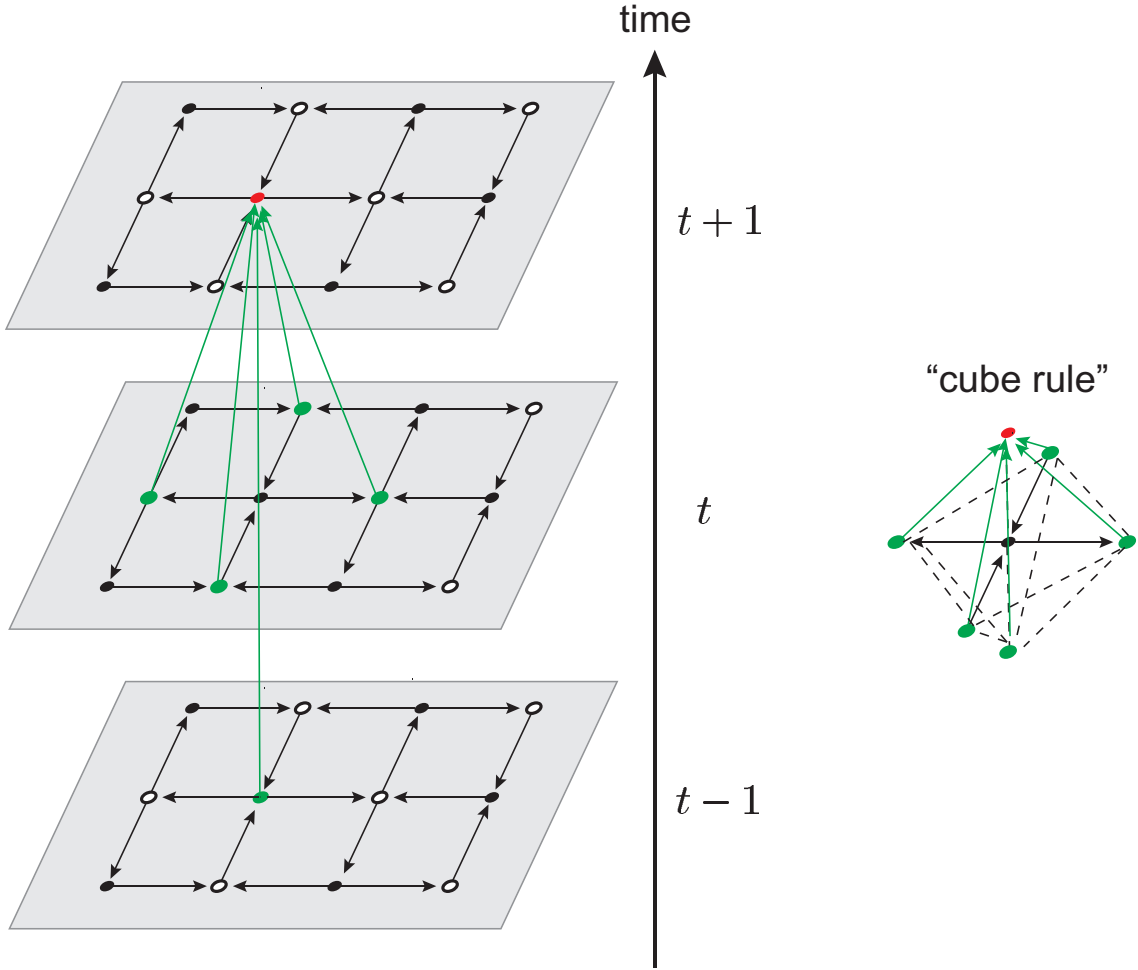}
\caption{The time evolution rule, for the case $G=A_3, G'=A_4$. The quiver of Fig.~\ref{fig.quiver} is placed on a fixed time slice,
where the time evolves in the vertical direction.
The time evolution rule (which we call the ``cube rule'' for $G=A_N$ or $A_N^{(1)}$ with $N>1$) is such that the dynamical variable at the red vertex at time
$t+1$ is determined by those at the green vertices, at times $t$ and $t-1$.
In other words, if we know the values of dynamical variables for all those on the vertices of a cube but one, then we obtain the remaining dynamical variable.
}
\label{fig.evolve}
\end{figure}

This equation means that $\sfy_m^a(t+1)$ is determined by 
a finite number of $\sfy$-variables at times $t$ and $t-1$. 
As shown in Fig.~\ref{fig.evolve}, for the case of $G=A_N$ the 
time evolution is determined by the ``octahedron rule'', 
namely the unknown dynamical variable at a vertex of the cube is determined once we know those variables at all other vertices of the cube.\footnote{Such a time evolution pattern for the classical case is called the octahedron recurrence \cite{RobbinsRumsey},
see also literature on discrete integrable systems, such as \cite{DoliwaSantini,AdlerBobenkoSuris}. This is also 
reminiscent of tensor networks.} 

Note that the fixed boundary condition (F) stipulates that the 
terms involving $y_{0}^a(t)$ and $y_{L+1}^a(t)$
do not appear on the right hand side of \eqref{eq.evolve}.

For the simply-laced case discussed here, the Dynkin diagram is bipartite and 
the dynamics of the $\sfy$'s for white vertices (``white dynamics'') and those for black vertices
(``black dynamics'') decouple.
For this reason we can choose to keep only one of them, and in fact
in the formulation of the discrete Liouville theory in \cite{Faddeev:1985gy,Faddeev:1993pe,Faddeev:1997hp,Faddeev:2000if,Faddeev:2002ms}
only one such copy is retained. However, such a decoupling of black and white dynamics is no longer present when $G$ is non-simply-laced, as we will comment on Sec.~\ref{subsec.non-simply-laced}.

Notice that we can easily show the the causality of the system:
\begin{align}
[y_{i,t}, y_{j,t'}] =0  \quad \textrm{if} \quad \textrm{dist}(i,j) > |t-t'| \;.
\label{causality}
\end{align}
where the distance $\textrm{dist}(i,j)$ is defined as the number of edges 
in the shortest path (graph geodesic) connecting two vertices $i$ and $j$.
Note that in this definition we disregard the orientation of the graph,
and use only the resulting unoriented graph.

\begin{figure}[htbp]
\centering\includegraphics[scale=0.6]{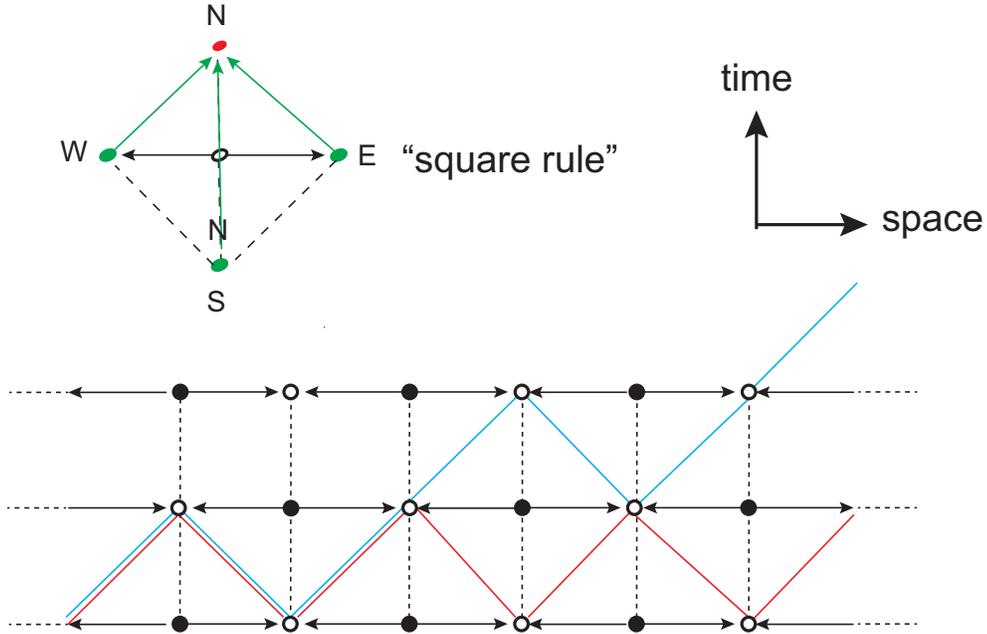}
\caption{For the case $G=A_1$, the time evolution rule is the ``square rule'', as shown above;
the dynamical variable at the red vertex is determined from those at the green vertices.
In this case, we can choose the discretized version of the Cauchy surface, for the 
time evolution for the dynamical variables associated with white vertices, as shown below
in the red line. Note that such a choice of the Cauchy surface is far from unique, and 
for example we can choose another surface as in blue.
}
\label{fig.Cauchy}
\end{figure}

The minimal set of $\sfy_m^a(t+1)$'s needed for the future/past time evolution is 
shown in Fig.~\ref{fig.Cauchy}. We can think of this as the discretization of the 
Cauchy surface.

\subsection{Motivation/Derivation}

\subsubsection{Classical Limit and Y- and T-System}

In order to motivate the time evolution rule \eqref{eq.evolve},
let us consider the classical limit $q\to 1$.
Then the variables $\sfy^a_m(t)$ mutually commute, which therefore we denote by 
$y^a_m(t)$.
The time evolution rule reduces to
\begin{align}
y_m^a(t+1) \, y_m^a(t-1) 
&=
\frac{
\prod_{b\ne a} (1+y_m^{b}(t))^{-C_{ab}}
}{
\prod_{n\ne m}(1+y_{n}^a(t)^{-1})^{-C'_{mn}} 
}\nonumber \\
&=
\frac{
\prod_{b\ne a} (1+y_m^{b}(t))^{-C_{ab}}
}{
(1+y_{m-1}^a(t)^{-1}) (1+ y_{m+1}^a(t)^{-1})
} 
\;,
\end{align}
where $C'$ is the Cartan matrix for $G'$.

This set of equations is known as the (classical) Y-system of 
type $(G, G')$ \cite{Ravanini:1992fi}.\footnote{For fixed boundary condition ($G'=A_L$), this is also known 
the Y-system of type $G$ at level $L$. See \cite{Kuniba:2010ir} for survey on Y- and T-system. It is worth mentioning that Y-system also appeared in  \cite{Gromov:2009tv,Alday:2010vh}.
}\footnote{In the literature on Y- and T-systems it is most common to denote the Y-system (T-system) in terms of capitalized letters $Y_m^a(t)$ ($T_m^a(t)$). In this paper we use an uncapitalized letters 
 $y_m^a(t)$ ($t_m^a(t)$), to better match with the cluster algebra notation.
}
For the case of the fixed boundary condition, we set
$y_{0}^a(t)^{-1}=y_{L+1}^a(t)^{-1}=0$ at the endpoints, just as in the quantum case.

The Y-system can be derived from a 
discretization of the Toda equation of the Hirota-type \cite{Hirota1}, 
known as the T-system. This reads
\begin{align}
t_m^a(t+1) t_m^a(t-1) 
&= 
\prod_{n\ne m} t_{n}^a(t)^{-C_{mn}}+\prod_{b\ne a} t_{m}^{b}(t)^{-C_{ab}}\nonumber \\
&= 
 t_{m-1}^a(t) t_{m+1}^a(t)+\prod_{b\ne a} t_{m}^{b}(t)^{-C_{ab}} \;,
 \label{classical_y}
\end{align}
where $ t_{m}^{r+1}(t)=1$. 

Let us first rescale the arguments by a factor of $\epsilon$, e.g.\
$t_m^a(t\pm 1)  \to \tau^a(x, t \pm\epsilon),
t_{m\pm 1}^a(t) \to \tau^a(x\pm \epsilon, t)$
where $x$ and $t$ are kept finite while $\epsilon$ is send to be zero in the continuum limit.
Expanding the resulting equation with respect to 
quadratic order in $\epsilon$, we obtain
\begin{align}
\frac{\partial^2 }{\partial u \partial v} \log \tau^a 
=\frac{1 }{(\tau^a)^2}\left[
\frac{\partial^2\tau^a}{\partial u \partial v}\tau^a
-  
\frac{\partial\tau^a}{\partial u}  \frac{\partial\tau^a}{\partial v}
\right]
 =\prod_{b} (\tau^{b})^{-C_{ab}} 
\;,
\end{align}
where we defined the light-cone coordinates $(u,v)$ by 
$u:=(t+x)/2, v:=(t-x)/2$. This is the Hirota bilinear form of the two-dimensional Toda equation.

There is one subtlety in the T-system, which is that it allows for some ``gauge ambiguity''.
Let us for example consider the case $G=A_{L-1}^{(1)}$, for which case the T-system reads
\begin{align}
t_m^a(t+1) t_m^a(t-1) 
&= 
 t_{m-1}^a(t) t_{m+1}^a(t)+  t_{m}^{a+1}(t) t_{m}^{a-1}(t)  \;,
\end{align}
This has an ambiguity of the form
\begin{align}
t_m^a(t)\to  \left[\prod_{s_1, s_2=\pm 1} f_{s_1, s_2}(t+s_1 m+s_2 a)\right]\,  t_m^a(t) \;.
\end{align}
While it is possible to first quantize and then divide by this gauge ambiguity, 
it is often more economical to first mod out by this gauge transformation and then quantize.
In this case, we should consider the gauge-invariant combination,
which leads to
\begin{align}
y_m^a(t)= \frac{t_m^{a+1}(t) t_m^{a-1}(t)}{t_{m+1}^a(t) \, t_{m-1}^a(t)} \;,
\end{align}
or more generally for simply-laced $G$
\begin{align}
y_m^a(t)= \frac{\prod_{b\ne a} t_m^{b}(t)}{t_{m+1}^a(t) \, t_{m-1}^a(t)} \;,
\end{align}
We can verify that this satisfies the classical Y-system, and thus coincides with 
the classical limit of the quantum variables $\sfy_m^a(t)$.

\subsubsection{Quantization \`{a} la Cluster Algebra}

Having explained the rule in the limit, let us now come to the quantum case.
Basically, what should be done is to promote the classical variables $y_m^a(t)$ into
non-commutative variables $\sfy_m^a(t)$ obeying \eqref{xCCR}. In particular, 
when we replace the classical $y$-system \eqref{classical_y} by its quantized version,
we need to include appropriate power of $q$ (which we do not see in the classical limit) 
in such a way that the resulting 
expression is consistent with the commutation relation \eqref{xCCR}.

For some simple cases we can play around with the expressions, and after some trial and errors we can arrive at the expression \eqref{eq.evolve}. However, we can also appeal to the 
more general mathematical theory of the so-called quantum cluster algebras, which is a quantization of the classical cluster algebras \cite{FominZelevinsky1}.\footnote{Our notation mostly follows those of \cite{Terashima:2013fg,Gang:2015wya,Gang:2015bwa}.}
This makes it possible to borrow some machineries developed for theory there.
In this paper we do not provide a detailed explanation of the quantum cluster algebra, and interested readers are referred to App.~\ref{sec.cluster} and e.g.\ to \cite{KashaevNakanishi,Gang:2015wya}.

In the cluster algebra we have two defining ingredients. 

First, we have a quiver $Q$, which determines the algebra at a fixed time slice (this corresponds to
Sec.~\ref{subsec.dynamical} and Sec.~\ref{subsec.commutation}.
For our case, the quiver is the $(G, G')$ quiver \eqref{QGG}, already shown in Fig.~\ref{fig.quiver}.
We associate dynamical variable $\sfy_i$, the so-called quantum $y$-variable\footnote{The
$y$-variable corresponds to the Y-system, whereas
T-system corresponds to the so-called cluster $x$-variables. }
 \cite{FockGoncharovEnsembles,FockGoncharovQuantumCluster}, to 
each vertex $i$ of the quiver $Q$.\footnote{If $q$ is root of
 unity (as discussed above to make the Hilbert space finite), then the quantum $y$-variable reduces to the cyclic cluster variable, as discussed in \cite{Ip:2014pva}.
} 

Second, we specify a sequence of operations called ``mutations'' (see \eqref{Qmutate} in Appendix), where each mutation 
 is labeled by a vertex $i$ of the quiver and is denoted by $\mu_i$.
Then a sequence of mutations is labeled by a sequence of vertices $\{ i_1, i_2, \ldots, \}$.
 
Such a sequence of mutations corresponds to a non-trivial time evolution (this is a counterpart of Sec.~\ref{subsec.evolution}).
It turns out that one step of the 
time evolutions of the discrete Toda theory (from time $t$ to $t+1$) corresponds to mutations at all the vertices of the quiver,
colored either black or white in Fig.~\ref{fig.quiver}.\footnote{Since none of the white vertices are connected with each other, the ordering of such mutations does not matter.}
Namely, if we define
\begin{align}
\mu_{\circ}:=\prod_{i: \circ} \mu_i  \;,\quad
\mu_{\bullet}:=\prod_{i: \bullet} \mu_i \;,
\label{mumu}
\end{align}
then the time evolution from time $t$ to $t+1$
corresponds to mutations $\mu_{\circ}$ or $\mu_{\bullet}$,
depending on whether $t$ is even or odd.
It turns out that the order of products in \eqref{mumu}
do not matter, since consecutive mutations at non-adjacent vertices is known to commute with each other and since the quiver is bipartite.

The time evolution rule \eqref{eq.evolve} in the quantum theory then follows from the 
transformation rules \eqref{quantum mutation} of quantum $y$-variables.


\bigskip

There is yet another advantage of the cluster-algebraic reformulation.
Namely, we can write a time-evolution operator $U_t$
\begin{align}
U_t\, \sfy^a_m(t') \, U_t^{-1} = \sfy^a_m(t+t') \;,
\label{Ut}
\end{align}
so that in the \Schrodinger picture the state
evolves by $U_t$:
\begin{align}
|\psi \rangle \to  |\psi(t) \rangle= U_t |\psi \rangle \;.
\end{align}
This is because such an operator for each mutation $\mu_i$, satisfying
\begin{align}
\mu_i : \quad \sfy_i \to U_t \, \sfy_i \, U_t^{-1} \;,
\end{align}
has already been constructed explicitly in the literature \cite{FockGoncharovQuantumCluster,KashaevNakanishi,Terashima:2013fg,Gang:2015wya}, see \eqref{mu_k}.
More concretely, such an operator $U_{i}$ 
can be written in terms of the quantum dilogarithm function (whose argument contains an operator $\sfy_i$), a linear operator mixing among $\sfy$'s, and a permutation operator. We can then define our time-evolution operator to be one of the following, depending
whether $t$ is even or odd:
\begin{align}
U_{\circ}=\prod_{i: \circ} U_i  \;,\quad
U_{\bullet}=\prod_{i: \bullet} U_i \;.
\end{align}
In this sense, the time evolution of the discrete Toda theory has already been solved.

For example, suppose that the spatial direction is periodic. 
We can choose an initial state $|\psi \rangle$ and a final state $|\tilde{\psi}(t) \rangle$, and compute the transition amplitude $|\langle \tilde{\psi}(t) |\psi \rangle|^2$
as an integral expressions \cite{Terashima:2013fg,Gang:2015wya}.
We then have the 
initial and final states at the past and future boundary circles of the annulus. 
By conformal transformation this can be mapped into a vacuum correlation function on a sphere.
Such a transition amplitude (and its trace) is known as the cluster partition function \cite{Terashima:2013fg,Gang:2015wya,Gang:2015bwa} (see also \cite{KashaevNakanishi}).\footnote{The discussion in these references are limited to the case where $G$ is simply-laced.}

If the final state coincides with the time evolution of initial state (namely if $|\tilde{\psi}(t) \rangle=|\psi(t) \rangle$), then the overlap amplitude $|\langle \psi(t) |\psi \rangle|^2$
is known as the survival amplitude (or Loschmidt echo).

For a typical thermal system, a small disturbance of the initial state by a local operator 
is expected to thermalize quickly and to be washed away, in time scale of the dissipation time, of order the inverse temperature.

The situation in our model seems to be very different, 
for the case of the fixed boundary condition (F). It turns out that the time evolution is periodic,
with period of order $L$ (which can be approximately regarded as inverse temperature of the system):
\begin{align}
\sfy^m_a(u+2(L+h_G^{\vee})) = \sfy^m_a(u) \;,
\label{periodicity}
\end{align}
where $h_G^{\vee}$ is the dual Coxeter number of $G$. This is known as the periodicity of the Y-system, as conjectured in \cite{Zamolodchikov:1991et,Kuniba:1992ev,Gliozzi:1994cs}
and proven later in \cite{Gliozzi:1995wq,Frenkel:1995vx,KellerPeriodicity} (including the non-simply-laced cases in \cite{IIKKN1,IIKKN2}), see \cite{Kuniba:2010ir} for more references.\footnote{Typically, periodicity of the Y-system is stated for the classical Y-system, however periodicity of the classical Y-system is actually equivalent with that for the quantum Y-system, as proven more generally in  \cite{FockGoncharovQuantumCluster}.
}\footnote{
One consequence \eqref{periodicity} is that product of the time-evolution operator 
$U_t$ from $t=1$ to $t=2(L+h_G^{\vee})$ is trivial. This gives rise to the 
quantum dilogarithm identity (see \cite{KashaevNakanishi} for more details).
}

While a quantum mechanical system with a discrete spectrum in general is known to show
a quantum recurrence phenomenon \cite{Bocchieri},
this case is very special since the recurrence time here grows linearly in the degrees of freedom, not in double exponentially.\footnote{Another difference from the general case is that the state here come back to
exactly the same state, whereas in general cases 
the state comes back only infinitely close to the original state.}  We might interpret this short-time revival as a signature of integrability of the model. We will comment more on this in Sec.~\ref{sec.chaos}.
Note that the period $2(L+h_G^{\vee})$ goes to infinity in the long-strip limit $L\to \infty$.
A similar periodicity is not known for the case of the periodic boundary condition (P).

\subsection{More General \texorpdfstring{$G$}{G}}\label{subsec.non-simply-laced}

Let us now comment on the case where $G$ is a more general algebra.
This includes three different generalizations;
\begin{itemize}\parskip=-2pt
\item A non-simply-laced finite-dimensional simple Lie algebra
\item A Lie superalgebra $SU(M|N)$
\item An infinite-dimensional tamely-laced Kac-Moody algebra
\end{itemize}

The basic story stays the same in all of these cases.
First, classical T- and Y-systems for these cases are already known in the literature \cite{Kuniba:2010ir},
which we can regard as the discretization of the the Toda equation.
Then these equations can be reformulated in the language of classical cluster algebras,
and by following the quantization procedure of quantum cluster algebra
we obtain the discretization of the  quantum Toda theory, as we wanted.

For the case of an infinite-dimensional Kac-Moody algebra, we need to impose a
technical condition that the Kac-Moody algebra is tamely-laced \cite{HernandezCoproduct,Kuniba:2009rh}.
Recall that a Kac-Moody algebra is defined from a generalized Cartan matrix $C$.
This Kac-Moody algebra is called {\it tamely-laced} \cite{HernandezCoproduct} if $C$ is symmetrizable and satisfies
\begin{align}
d_i=-C_{ji}=1 \, \quad \textrm{if} \quad C_{ij}<-1\;,
\label{tame_laced}
\end{align}
where $D=\textrm{diag}(d_1, \ldots, d_r)$ is a diagonal matrix symmetrizing $C$.
This includes most of the affine Lie algebras, except for $A_1^{(1)}$ and $A_{2N}^{(2)}$.

In order to highlight some subtleties in these generalizations, 
let us here take the simplest non-simply-laced example, namely $B_2$ (see \cite{IIKKN1,IIKKN2} for more details on the non-simply-laced cases).

\begin{figure}[htbp]
\centering\includegraphics[scale=0.6]{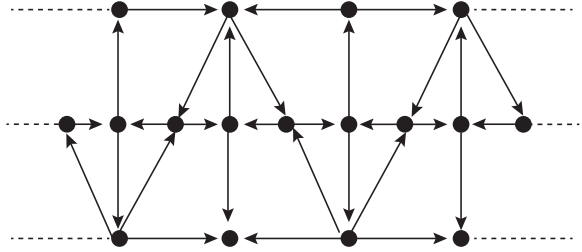}
\caption{Quiver $Q$ for $G=B_2$. For fixed boundary condition (F) of length $L=5$,
we can delete all the dotted lines and the figure gives the complete quiver. Note that this quiver is not bipartite.}
\label{fig.B2quiver}
\end{figure}

The quiver for the $G=B_2$ with a fixed boundary condition ($G'=A_L$) is shown in Fig.~\ref{fig.B2quiver}. 
This quiver is not bipartite, and the vertices are labeled by $(a,m)$,
with $a=1,2$ and $m=1, \ldots, t_a L-1$, with $t_1=1, t_2=2$.
Such a difference arises since $\alpha_1$ ($\alpha_2$) is a long (short) root.

The time evolution (quantum Y-system) is given by
the quantization of the classical Y-system known in the literature \cite{IIKKN2}:
\begin{align}
\begin{split}
&\sfy_m^{1}(t+1)\, \sfy_m^{1}(t-1) = \frac{
(1+q\, \sfy_{2m-1}^{2}(t))
(1+q\, \sfy_{2m+1}^{2}(t))
(1+q\, \sfy_{2m}^{2}(t-\frac{1}{2}))
(1+q\, \sfy_{2m}^{2}(t+\frac{1}{2}))
}{
(1+q\, \sfy_{m-1}^{1}(t)^{-1})
(1+q\, \sfy_{m+1}^{1}(t)^{-1})
}\;, \\
&\sfy_{2m}^{2}\left(t+\frac{1}{2}\right) \sfy_{2m}^{2}\left(t-\frac{1}{2}\right) = \frac{
1+q\, \sfy_{m}^{1}(t)
}{
(1+q\, \sfy_{2m-1}^{2}(t)^{-1})
(1+q\, \sfy_{2m+1}^{2}(t)^{-1})
} \;,\\
&\sfy_{2m+1}^{1}\left(t+\frac{1}{2}\right) \sfy_{2m+1}^{1}\left(t-\frac{1}{2}\right)= \frac{
1
}{
(1+q\, \sfy_{2m}^{2}(t)^{-1})
(1+q\, \sfy_{2m+1}^{2}(t)^{-1})
}
\;.
 \label{B2_c}
 \end{split}
\end{align}
Note that for $\sfy^a_m(t)$-variable corresponding to the short root ($a=2$),
the time-evolution step is half that for the long root ($a=1$). However, it is also the case that the range of $m$ is doubled, and hence the spatial spacing is also reduced by half. This means that 
that we still have the causality \eqref{causality} with the same velocity of light, as long as 
we modify the definition of distance accordingly; such a definition of the distance is natural 
when we draw the quiver on a two-dimensional plane, as down in Fig.~\ref{fig.B2quiver}.

\subsection{Exchange Symmetry}

An unexpected feature of our construction is that the 
$(G, G')$ quiver in Fig.~\ref{fig.quiver} has an obvious symmetry exchanging $G$ and $G'$. This symmetry is also observed in our time evolution rules.

Since $G'$ is taken to be either $A_L$ or $A_{L-1}^{(1)}$, 
this symmetry is present if
\begin{itemize}\parskip=-2pt
\item Fixed boundary condition (F) and $G=A_L$
\item Periodic boundary condition (P) and $G=A_{L-1}^{(1)}$
\end{itemize}

In either case, the symmetry exchanges $N$ and $L$, and the corresponding indices $a$ and $m$:\footnote{T-system is mapped as $t_m^a(t) \to t_a^m(t)$.}
\begin{align}
N  \to L\;, \quad
\sfy_m^a(t) \to \sfy_a^m(t)^{-1} \;.
\label{symmetry}
\end{align}
This symmetry is known as the level-rank duality of the Y- (and T-) system, as noted in the context of the RSOS model \cite{Bazhanov:1989yk,Kuniba:1990zh}.\footnote{Such a symmetry
has been discussed in a rather different context in \cite{Cecotti:2010fi,Cecotti:2014zga,Heckman:2012jh}.}

While this symmetry is trivial from the standpoint of the Y-system, 
this exchange symmetry \eqref{symmetry} is rather surprising from the standpoint of the discrete Toda theory, since 
the role of the symmetries $G$ and $G'$ are completely different--$L$ is the size of the spatial direction, while $N$ is the rank of the symmetry algebra of the theory, and a-priori there is nothing to indicate the symmetry between the two.
It is also interesting that the massive deformation of a CFT, replacing $G=A_N$ by $A_N^{(1)}$, is translated into the change of the boundary condition, from fixed to periodic.

Notice that for a general choice of $G$ (e.g.\ the non-simply-laced $G$) the choice $(G, A_L)$ now breaks the $L\leftrightarrow N$ exchange symmetry mentioned above. This might motivate us to consider a more general $(G, G')$ theory,
where $G$ specifies the type of the Toda lattice and $G'$ the spatial direction.
For exceptional $G'$ the length of the Dynkin diagram is bounded by above, and hence we have trouble taking the continuum limit. 
Moreover, if we wish to obtain a periodic spatial directions then $A_L^{(1)}$, which has a circular affine Dynkin diagram, is the only possible option, at least when $G$ is either a finite or affine Lie algebra.

\section{Quantum Chaos}\label{sec.chaos}

\subsection{Bounds on Quantum Chaos}

As stated in introduction, discrete Liouville theory was recently proposed to be
maximal chaotic \cite{Turiaci:2016cvo}.

Recall that in classical systems the chaos refers to the sensitivity of the system
to the initial condition; small perturbations of the initial conditions
grows exponentially in time. In quantum systems it is more subtle to 
define chaos. The recent proposal is to use the 
out-of-time correlator of the form $\langle W(t) V W(t) V \rangle_{\beta}$,
where $V$ and $W$ are generic operators of the theory,\footnote{The operators 
$V$ and $W$ need to be smeared in the time direction, to avoid singularities.
In practice this is naturally incorporated in the $i\epsilon$-prescription for the 
analytic continuation from Euclidean to Lorentzian signature \cite{Roberts:2014ifa}.}
and the subindex $\beta$ refers to the evaluation at a thermal state with temperature $\beta$.
The quantum chaos is then characterized by the 
exponential growth of this correlator, 
as a function of time $t$:
\begin{align}
\langle W(t) V(t=0) W(t) V(t=0) \rangle_{\beta} \sim
\textrm{(const.)}- e^{\lambda_L(t-t_*)} \;.
\end{align}
Here $t_*$ is some time scale after which the exponential growth begins.
Also, the exponent $\lambda_L$  is called the Lyapunov exponent;
quantum chaos is characterized by $\lambda_L\ne 0$, and the larger the value of $\lambda_L$
the more chaotic the system is.

In classical system the value of the Lyapunov exponent can be 
arbitrary large. However, in quantum systems there exists a conjectured bound \cite{Maldacena:2015waa}\footnote{This assumes some hierarchy between dissipation time and scrambling time.}
\begin{align}
\lambda_L \le \frac{2\pi }{ \beta} \;,
\label{MSS_bound}
\end{align}
where $\beta$ is the inverse temperature and we have set $k_B=\hbar=1$.

For us, the interesting fact is that the quantum discrete Liouville theory saturates this bound,
so that we have $\lambda_L=2\pi /\beta$ \cite{Turiaci:2016cvo}.

\subsection{Quantum Chaos in Discrete Toda Theory}

Since our model is a natural generalization of the discrete Liouville theory, it is natural to ask if our model adds anything to these discussions. In this section we take $G$ to be a simple finite-dimensional Lie algebra,
in particular $G=A_{N-1}$, hence the Toda theory in the continuum is conformal.

The most direct method to tackle this problem is to 
evaluate the out-of-time correlator explicitly in our model;
we can try to take $W(t)$ and $V$ to be for example
$W=\sfy_m^a(t)$ and $V=\sfy_n^b(t=0)$. 
Here the inverse temperature $\beta$ can be identified with the length $L$ of the spatial direction.
Such a computation seems to be involved, and has not been done, even in the simplest case of 
$G=A_1$.

Instead let us here appeal to the fact that our model reduces to the Toda field theory in the continuum. Since we expect the Lyapunov exponent to be a characterization of the effective theory and to be UV-insensitive, we expect that the exponent for the discrete and continuum theories coincide.

The continuum theory, {\it i.e.} the Toda theory, is a two-dimensional CFT with $W_N$ symmetry,
and by taking the central charge to be large we expect that gravity is semiclassical in the 
holographic dual (recall that the Newton constant in the bulk is inversely proportional to the central charge).\footnote{For existence of semiclassical holographic dual, we also need to take into account the sparseness of the spectrum 
for the semiclassical holographic dual \cite{Heemskerk:2009pn,Hartman:2014oaa}.
} Such a bulk theory is known to be the $SL(N, \mathbb{R})$ Chern-Simons theory,
which contains particles with spin greater than $2$.

In this case, we can argue that the
$W_N$ conformal block for the identity operator
contributes to the Lyapunov exponent as (see \cite{Perlmutter:2016pkf},
which builds on the discussion for the $N=2$ case \cite{Roberts:2014ifa})
\begin{align}
\lambda_L= \frac{2\pi (N-1)}{\beta} \;.
\label{lambda_N}
\end{align}
One might therefore conclude that for $N>2$ this result violates the bound \eqref{MSS_bound}, and hence
the quantum discrete Toda theory is inconsistent, at least in the continuum limit.
However this is {\it not} correct---the derivation of \eqref{lambda_N} assumes the
dominance of the vacuum identity block, which does not hold in the quantum Toda theory.
That quantum Toda theory is consistent is far from trivial, since its bulk dual, namely $SL(N, \mathbb{R})$ Chern-Simons theory, contains a finite number of higher spin particles, which in general is known to violate causality (after suitable coupling to matters) \cite{Camanho:2014apa}.

It is therefore an important problem to compute the value of the Lyapunov exponent for 
quantum Toda theory, both for discrete and continuum cases. Note that one existent argument for the value of the Lyapunov exponent for the $N=2$ case \cite[section 2]{Turiaci:2016cvo} relies on some results in Liouville theory \cite{Teschner:1995dt,Balog:1997zz,Dorn:2006ys}, whose Toda ($N>2$) counterpart seems to be unknown in the literature.

We conjecture that the Lyapunov exponent is positive ($\lambda_L>0$) and hence is chaotic, 
for all values of $N\ge 2$.

For better understanding of our theory, one possibility is to consider the large $N$ limit, so that 
we have at we have an infinite number of higher spin particles in the bulk, where the apparent discrepancy between \eqref{MSS_bound} and \eqref{lambda_N} is sharpest.
The ``extra dimension'' has 
width $N$, hence decompactifies in the large $N$ limit. 
This is a version of the 
dimensional oxidation, where the two-dimensional lattice is turned into
a three-dimensional lattice.\footnote{This depends on the order of the two limits;
the large $N$ limit and the continuum limit. Most naively, we should first take a continuum limit, and then take the large $N$ limit.
However, that will give a diverging contribution to the Lyapunov exponent from the vacuum block (see \eqref{lambda_N}), signaling the need for resummation. This might motivate taking the large $N$ limit first.
This is somewhat reminiscent of the situation in \cite{Roberts:2014ifa}, where we first need to resum the global $SL(2, \mathbb{R})$
block into the Virasoro conformal block {\it before} taking the Regge limit.
}
Notice that thanks to the exchange symmetry between the spatial length $L$ and rank $N$,
the decompactified dimension is on equal footing with the spatial direction.

Such a large $N$ limit should be compared with the case of the two-dimensional CFT with the $W_{N=\infty}[\lambda]$ symmetry, which is dual to the Vasiliev theory  \cite{Vasiliev:1990en,Vasiliev:1999ba} in $\textrm{AdS}_3$ theory. 
This theory has a vanishing Lyapunov exponent ($\lambda_L=0$), and hence is not chaotic \cite{Perlmutter:2016pkf}.
What happens there, at least schematically, is that we have the re-sum the infinite series representing the 
infinite spins, and the result has a effective spin not larger than $2$, making the theory consistent.
One possible reason behind such a miraculous resummation is the integrability of the $W_{\infty}[\lambda]$ CFT. 
In this respect one should keep in mind that Liouville/Toda theory is also integrable. However, integrability in itself does not necessarily guarantee that the system is non-chaotic $\lambda_L =0$---many of the integrable charges are non-local, and the time evolution could happen in the basis where the charges are not conserved. One could also turn on a small non-integrable deformation of the system, to ensure that the system is chaotic \cite{Turiaci:2016cvo}.

\section{Relation with Higher \Teichmuller Theory}\label{sec.Teichmuller}

In this section, let us discuss the relation of our discrete model with the higher \Teichmuller theory.
The higher \Teichmuller theory in question will be defined on an annulus, 
and we choose the periodic boundary condition for the discrete Toda theory\footnote{Essentially the same argument can be repeated for the higher \Teichmuller theory on a strip
and the discrete Toda theory with fixed spatial boundary condition. However, one subtlety in this case is that we need to modify the definition of the Dehn twist on the boundary of the strip.
}
depending on the fixed or periodic boundary condition. 

\subsection{\texorpdfstring{$G=A_1$}{G=A1}}

The reference \cite{Faddeev:2002ms} pointed out the 
equivalence between discrete Liouville theory and the 
\Teichmuller theory on annulus. It was also pointed out that
the evolution operator $U_t$ (recall \eqref{Ut}) in the former coincides with  
the geometrical Dehn-twist operator of the latter.

The argument of \cite{Faddeev:2002ms} required some complicated computations
involving quantum dilogarithm functions. However,
from a modern perspective there is no need to go through such complicated computations, to establish the equivalence mentioned above---this equivalence follows from the simple observation that (a) the discrete Liouville theory with periodic spatial boundary condition and (b) \Teichmuller theory on annulus, are both described by the same quantum 
cluster algebra datum, namely by
the same quiver and the mutation sequence.

\begin{figure}[htbp]
\centering{\includegraphics[scale=0.6]{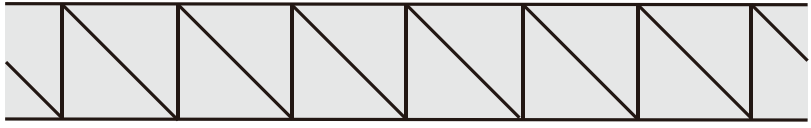}}
\caption{An ideal triangulation of the annulus in $A_1$ \Teichmuller theory.}
\label{fig.triangulation}
\end{figure}

To explain the cluster algebra structure for the quantum Liouville theory on annulus,
let us first consider a triangulation of the annulus, as in Fig.~\ref{fig.triangulation}.\footnote{In
\Teichmuller theory, we are supposed to consider an ideal triangulation, namely a triangulation where all the vertices are located on the punctures (or marked points on the boundary) of the surface. This means we actually have an annulus with $2L$ marked points, with $L$ each for the upper and lower circular boundaries of the annulus.
}

\begin{figure}[htbp]
\centering{\includegraphics[scale=0.6]{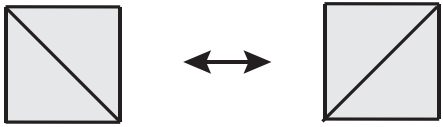}}
\caption{A flip, an exchange of the diagonal in a square, maps one triangulation to another.
The square here is meant to be a small part of a larger triangulation (a triangulation of an annulus in this case).}
\label{fig.flip}
\end{figure}

\begin{figure}[htbp]
\centering{\includegraphics[scale=0.6]{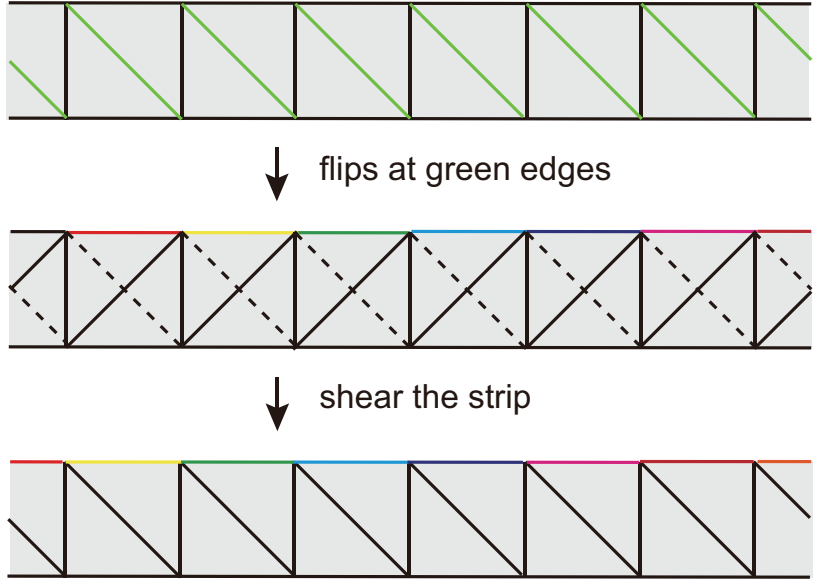}}
\caption{By applying multiple flips, we obtain another triangulation
of the annulus, as shown in the middle. By shifting the relative positions of the 
two boundary circles, we then obtain the picture below, 
which is pictorially the same as the figure we started with.
Note however the positions edge on the upper boundary circle are shifted
(as represented by colored intervals). This is known as the Dehn twist
(along a circle parallel to boundary circles).}
\label{fig.Dehn_twist}
\end{figure}

Given a triangulation of the annulus, we can change the triangulation, by 
applying the operation of Fig.~\ref{fig.flip} (call a flip) to one of the squares. By repeating this flip
as in Fig.~\ref{fig.Dehn_twist}, and then changing the relative positions of the 
two boundaries of the annulus, we can realize the  so-called Dehn twist, 
as applied to the triangulation (Fig.~\ref{fig.Dehn_twist}).

\begin{figure}[htbp]
\centering{\includegraphics[scale=0.6]{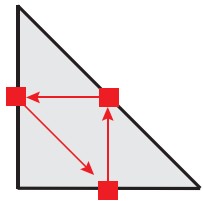}}
\caption{The quiver for a triangle, in the $A_1$ \Teichmuller theory. All the three vertices in this figure are shown as square, and hence is non-dynamical.}
\label{fig.A1quiver}
\end{figure}

In order to make contact with this geometrical picture with the more algebraic setup in cluster algebras, let us first associate a quiver of Fig.~\ref{fig.A1quiver}
to each triangle of the triangulation. Here the quiver has three vertices, each of which is shown as a square box. This means that the corresponding vertices (and the quantum cluster $y$-variables $\sfy_i$ associated with them) are non-dynamical (``frozen'' in the terminology of cluster algebras).

\begin{figure}[htbp]
\centering{\includegraphics[scale=0.6]{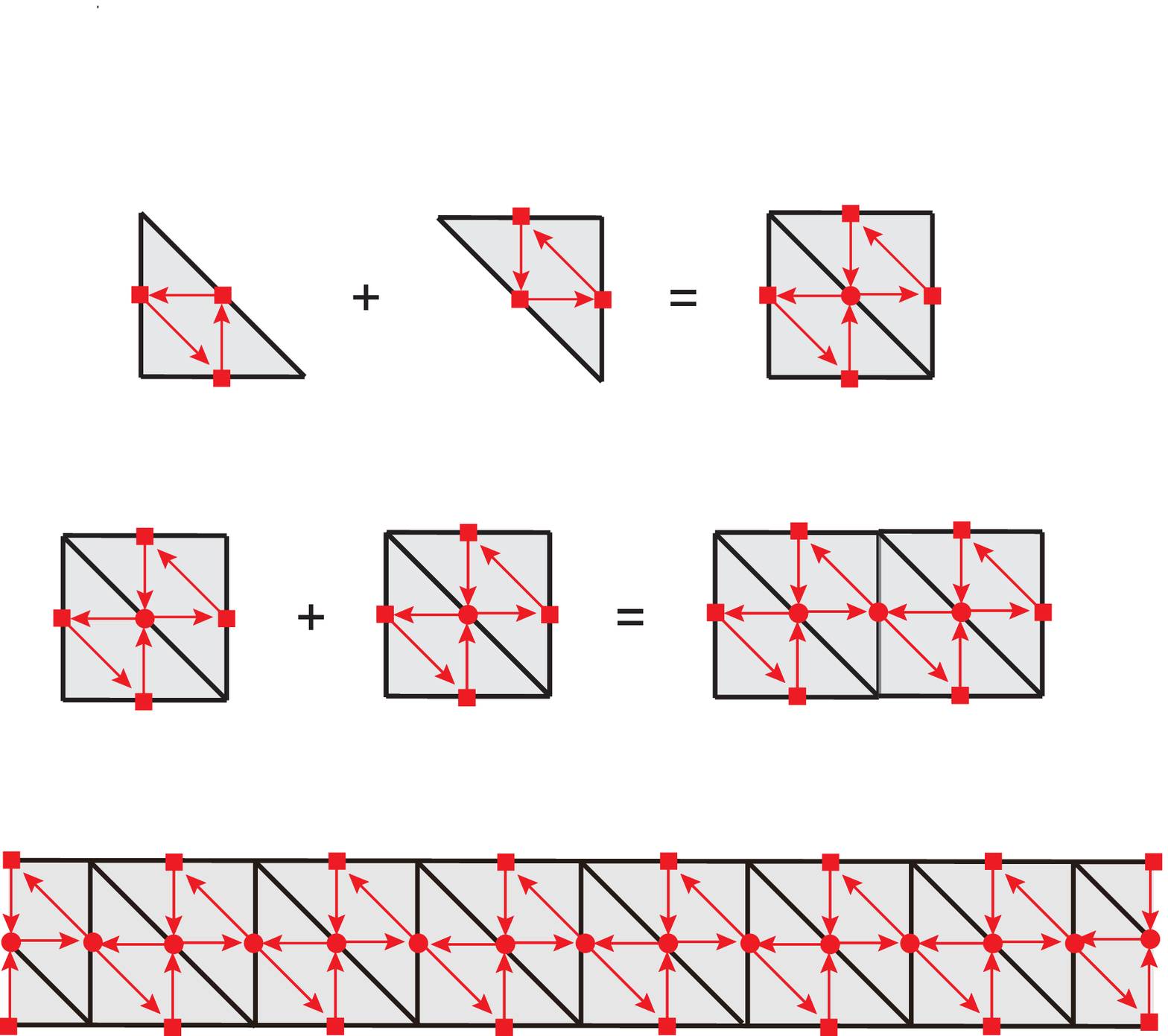}}
\caption{When we glue triangles, we glue the associated quivers inscribed on the triangles.
In this process, the vertices associated with glued edges are turned dynamical, and promoted from squares into circles. By repeating this procedure you obtain the 
quiver for the annulus, as shown in the bottom figure.}
\label{fig.A1_quiverglue}
\end{figure}

Now in the triangulation of an annulus  the triangles are glued together along their edges.
Whenever two triangles are glued together, we first concatenate the quivers by identifying the quiver vertices associated with the glued edge, and then promote that vertex (and the associated  quantum cluster $y$-variables $\sfy_i$) to be dynamical. We denote such a dynamical vertex by a circle in Fig.~\ref{fig.A1_quiverglue}. By repeating this procedure you obtain the 
quiver for the annulus, as shown in the bottom of Fig.~\ref{fig.A1_quiverglue}.\footnote{Such a construction can be sort of as an open analog of the Gaiotto's construction for gauging \cite{Gaiotto:2009we}, as emphasized in \cite{Xie:2012mr}}

Now the first non-trivial observation is that the resulting quiver for the annulus, 
once we disregard the non-dynamical (squared) vertices and the edges beginning/ending on them,
coincides with the $(G=A_1, G'=A_L^{(1)})$ quiver introduced in \eqref{QGG}.

\begin{figure}[htbp]
\centering{\includegraphics[scale=0.6]{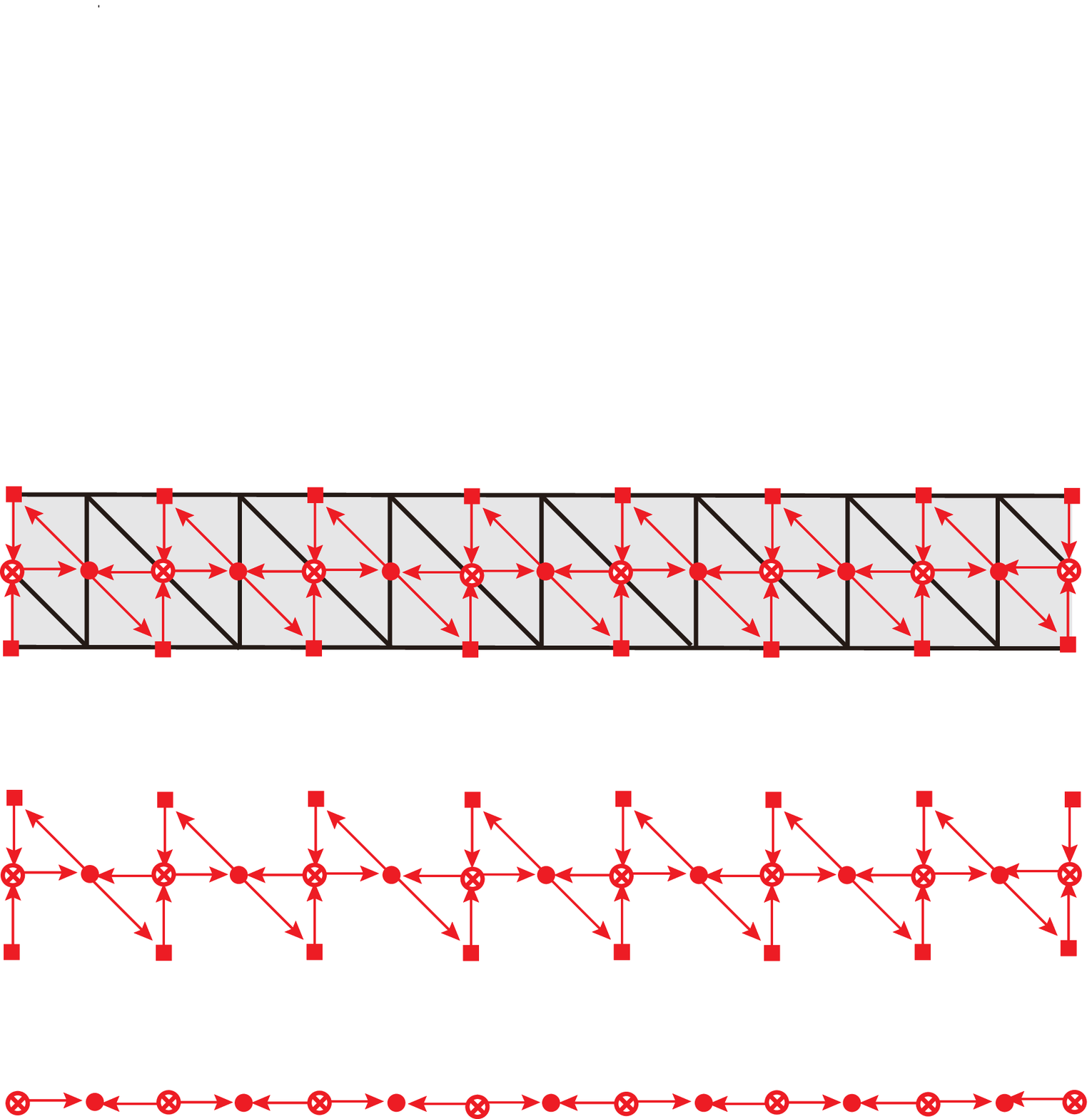}}
\caption{The Dehn twist  Fig.~\ref{fig.Dehn_twist} amounts to 
mutations at all the even/odd vertices. This matches the mutations rules for the discrete Toda theory in \eqref{mumu}. Note that for this discussion we disregard the non-dynamical (square) vertices, as shown in the bottom figure.}
\label{fig.A1_translate_2}
\end{figure}

We can moreover match the time evolutions, namely the mutation sequence of the quiver.
In the gluing rule of Fig.~\ref{fig.A1_quiverglue}, the flip of the triangulation (Fig.~\ref{fig.triangulation}) is turned into a change of the quiver as shown in Fig.~\ref{fig.A1_translate}. This is nothing but a mutation of the quiver (see Appendix), for a vertex on the flipped edge (as represented as a crossed vertex in the figure). Once we establish this, 
we can translate the flips for a realization of the Dehn twist  (Fig.~\ref{fig.Dehn_twist})
into a sequence of mutations, as in Fig.~\ref{fig.A1_translate_2}. The result is to 
mutate all the even (or odd) vertices. This coincides with our previous discussion (see explanation around \eqref{mumu}). This establishes what we wanted to show. Notice that for this purpose graphical/combinatorial manipulations are enough, and no complicated computations are necessary.

\begin{figure}[htbp]
\centering{\includegraphics[scale=0.6]{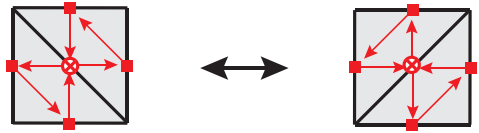}}
\caption{A flip of an ideal triangulation is translated into a mutation of the quiver, at a vertex located on the flipped edge.}
\label{fig.A1_translate}
\end{figure}

That there exists a correspondence between the quantum Liouville
theory and quantum \Teichmuller theory is in itself not surprising.
This is because classical Liouville theory originated in the study of the uniformization of a
Riemann surface, which deals with the \Teichmuller space. Such an equivalence was 
conjectured to persist at the quantum level \cite{Verlinde:1989ua}, which equivalence was later proven in \cite{Teschner:2005bz,Vartanov:2013ima}.

However, what is shown here is more dramatic, namely
we have a direct relation between quantum {\it discrete} Liouville
theory and quantum \Teichmuller theory in the {\it continuum}.
This is a rare example where ``a discretization of a theory 
reproduces the original theory''.
As we will see next, it turns out that this is a special feature of the $A_1$ case, and 
does not really hold for the $A_N$ case.


\subsection{\texorpdfstring{$G=A_N$}{G=AN}}

Let us now come to the case of $G=A_N$. The generalization of the quantum \Teichmuller theory for this is the higher \Teichmuller theory of \cite{FockGoncharovHigher}.

\begin{figure}[htbp]
\centering\includegraphics[scale=0.6]{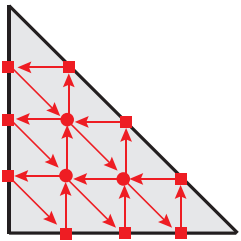}
\caption{In $A_N$ higher \Teichmuller theory, we associate a quiver shown in this figure for each ideal triangulation. The vertices on the boundary edges are non-dynamical (depicted as squares), whereas other vertices in the interior of the triangle are dynamical depicted as circles).}
\label{fig.FGquiver}
\end{figure}

The part about the triangulation of the annulus, as shown in Figs~\ref{fig.triangulation}, \ref{fig.flip} and \ref{fig.Dehn_twist}, stay the same. The difference comes for the rule for the quiver (Fig.~\ref{fig.A1quiver}), which for the $A_N$ case is given in Fig.~\ref{fig.FGquiver}.
This quiver has $3(N-1)$ frozen vertices on the boundary edge, which are regarded as non-dynamical. Note the quiver also has $(N-1)(N-2)/2$ dynamical vertices in the interior of the triangle.

The gluing rule, previously shown in Fig.~\ref{fig.A1_quiverglue}, stay essentially the same.
The only difference is that whenever we glue an edge of the two triangles $N-1$ vertices are turned dynamical for general $N$.

\begin{figure}[htbp]
\centering\includegraphics[scale=0.6]{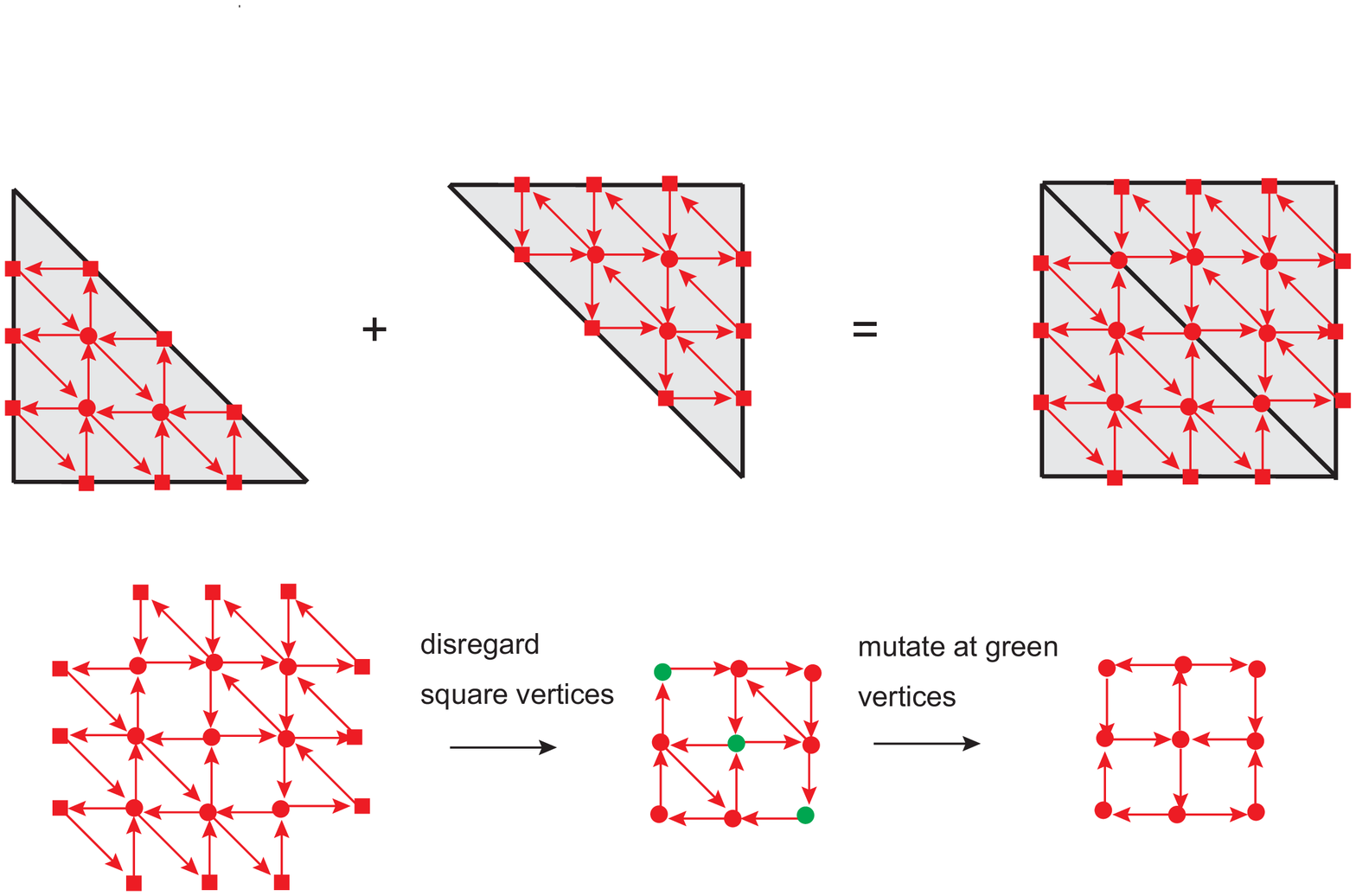}
\caption{The quiver for the  $A_N$ higher \Teichmuller theory for a square, with two triangles glued together. By disregarding the non-dynamical vertices, and mutating at the green vertices,
we obtain the quiver shown in Fig.~\ref{fig.quiver}. It turns out that this quiver is mutation-equivalent with the 
$(A_N, A_N)$ quiver of Fig.~\ref{fig.quiver}.
}
\label{fig.A3glue}
\end{figure}

Let us first glue two triangles, to obtain a square. In this case, we obtain a quiver in Fig.~\ref{fig.A3glue}, which looks very different from the $(A_N, A_N)$ quiver in Fig.~\ref{fig.quiver}.  One might therefore conclude that the relation between the two theories is lost completely.

It turns out, however, that there is a sequence of mutations relating the two (see Fig.~\ref{fig.A3glue}), and hence
the two quantizations are simply related by some unitary transformation $U$:
\begin{align}
\sfy_i^{\textrm{dT}}(t) = U^{-1} \,\sfy_i^{\textrm{hT}}(t) \, U 
\end{align}
where $\sfy_i^{\textrm{dT}}(t)$ ($\sfy_i^{\textrm{hT}}(t)$) denotes the $\sfy$ variables of the 
discrete Toda (higher Teichm\"{u}ller) theory.\footnote{One should not that such a unitary transformation is not local on the lattice, since mutations mixes variables on neighboring vertices of the quiver. This means that the 
causality \eqref{causality}, which holds for $\sfy_i^{\textrm{dT}}(t)$, does not hold for $\sfy_i^{\textrm{hT}}(t)$, for the case $G=A_{N}$ with $N>1$. 
}

\begin{figure}[htbp]
\centering\includegraphics[scale=0.6]{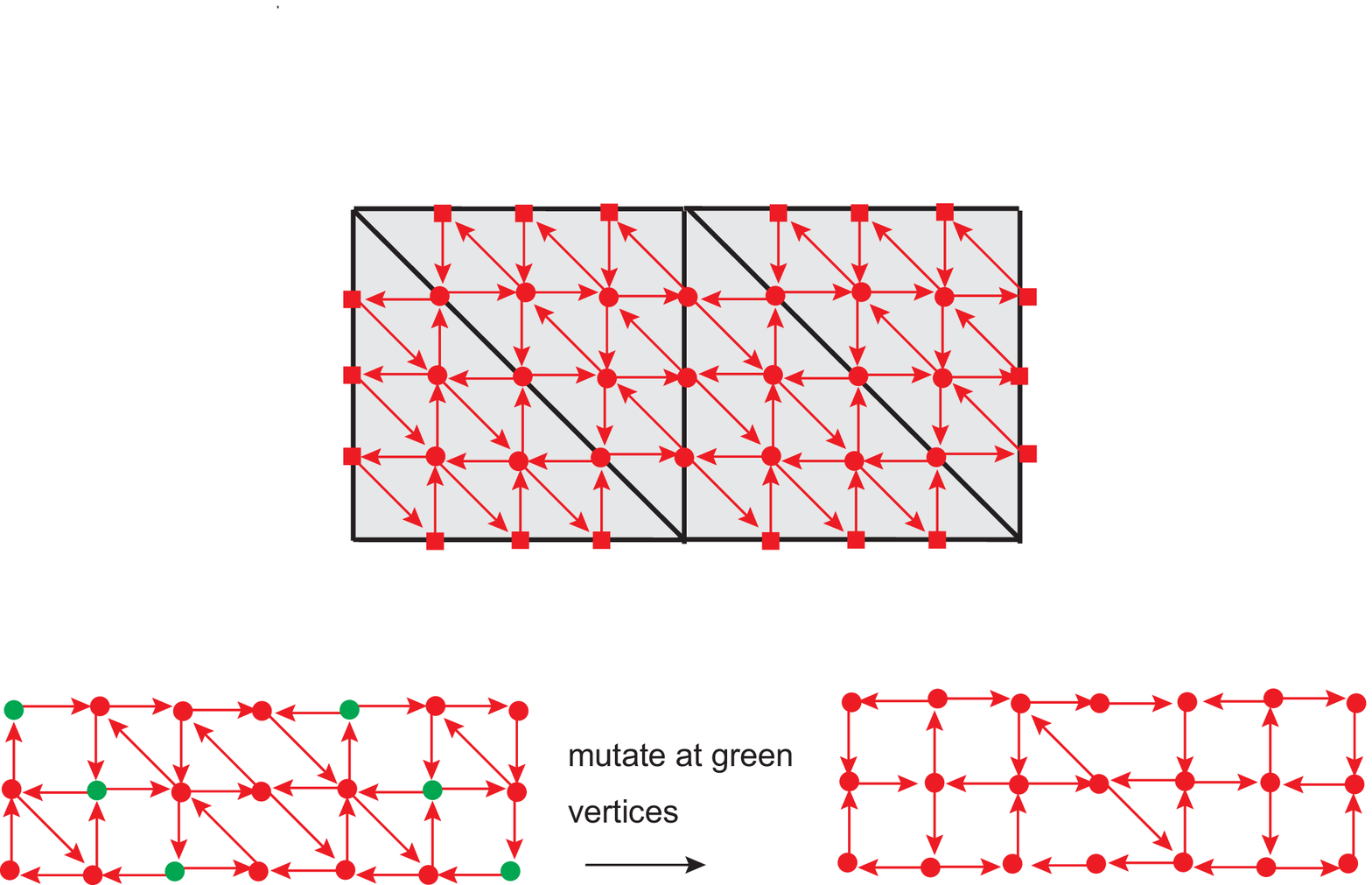}
\caption{The quiver for the  $A_N$ higher \Teichmuller theory for two square, with four triangles glued together (above). We can again disregard the non-dynamical vertices, and obtain the quiver as in bottom left. After some mutations, however, the resulting quiver (bottom right) is still different from the quiver in Fig.~\ref{fig.quiver}, even if we take $N=3, L=7$, as required from the match of the number of quiver vertices.
}
\label{fig.comparison2}
\end{figure}

While this is encouraging, such a nice story ceases to exist once we begin to glue two squares (hence four triangles). Indeed, the resulting quiver as required by the quantum \Teichmuller theory
is shown in Fig.\ref{fig.comparison2}, and even after mutations still is different 
from the quiver \eqref{QGG} of Fig.~\ref{fig.quiver}.
The basic reason for this is that when we glue two squares we turn the non-dynamical vertices into dynamical vertices, and the structures of the non-dynamical vertices are 
different between the two theories, already for a square (two triangles).
This means that the direct relation between discrete $A_N$ Toda theory and continuum $A_N$ higher \Teichmuller theory does not hold, for $N>2$.

This is not a surprising statement, since as we discussed before there is no a-priori reasoning to guarantee an equivalence between the two. Nevertheless it would be interesting to explore further if there is anything we can extract by the similarities of the two subjects, even in the case of $N>2$. For example, the discrepancy between the two comes from gluing edges, which are
locate on one-dimensional edges and hence would be suppressed compared with those degrees of freedom on the interior of the triangles, in the large $N$ limit. This could be another indication that large $N$ limit has some special properties.

\section{Summary and Discussion}\label{sec.conclusion}

In this paper, we formulated the discrete Toda theory from the quantum Y-system
associated with the quiver $Q=G\Box G'$. Here $G$ is a symmetry algebra of the theory,
which can be a finite-dimensional semisimple Lie algebra or an infinite-dimensional tamely-laced Kac-Moody algebra. Another algebra $G'$ is either $A_L$ or $A_{L-1}^{(1)}$, depending on the 
spatial boundary conditions. Our formulation naturally generalizes the quantum $A_1$ Liouville theory in the literature, however as we discussed in Sec.~\ref{sec.Teichmuller} the direction relation with the 
$A_N$ higher \Teichmuller theory on annulus seems to be lost for $N>2$. We also commented on possible implications to chaos.

Let us here comment on some more open questions which are not touched in the main text.

First, it would be interesting to see if the discretized model in this paper helps to 
solve the Toda field theory in the continuum limit, for example to 
compute the three-point structure constants and the four-point conformal blocks.
While the details of the computation might be involved, our discretized/regularized model is `solved' already, and might give rise to a systematic method to solve the continuum Toda theory.

As commented in the main text, our model corresponds to particular examples of
the quantum cluster algebra, and the discussion naturally generalizes to more general
choices of quivers and mutation sequences ({\it cf.}\ \cite{Inoue:2010si,Nakanishi:2010uv}). It would be interesting to identity (if any)
the two-dimensional CFT in the continuum limit, and compute their Lyapunov exponents.
The hope is that this generalization provides a rich landscape of discrete quantum mechanical systems to explore quantum chaos.

In Sec.~\ref{sec.Teichmuller} we worked on the relation between discrete Toda theory and the higher \Teichmuller theory, for the $A_N$ case. While the conclusion was negative overall, we also obtained a positive result, namely that some structures of the higher \Teichmuller theory (namely the dynamical part of the quiver for a square) can be extracted from the corresponding Y-system.
It would be interesting to verify this for more general case $G$.
Note that the general $G$ version of the higher \Teichmuller theory is being developed only recently \cite{Le1,Zickert16}. 

In our model, we discussed two types of spatial boundary conditions, fixed and periodic. This is motivated partly by simplicity, and partly by those often used in the literature of classical Y-system.
However, we have not tried to find exhaustive list of boundary conditions consistent with 
integrability of the model. Note that the classical boundary conditions preserving integrability 
are highly constrained in the continuum in the continuum affine Toda theories \cite{Bowcock:1995vp,Delius:1998he}. 
In this respect, 
taking the general algebra $G'$ in the $(G, G')$-quiver (say for $G'=B_n$ or $A_{2k-1}^{(2)}$) could realize some interesting integrable boundary conditions.


\section*{Acknowledgments}
The author would like to thank Richard Eager, Simeon Hellerman, Ivan Ip, Atsuo Kuniba and Herman Verlinde for illuminating discussion and correspondence.
The author benefited from his presentation on discrete Liouville theory in May 2016 at IAS, 
and he would like to thank the audience for feedback.
This research is supported by WPI program (MEXT, Japan), by JSPS Program for Advancing Strategic International Networks to Accelerate the Circulation of Talented Researchers, by JSPS KAKENHI Grant No.\ 15K17634, and by JSPS-NRF Joint Research Project.


\appendix

\section{Continuum Theory}\label{subsec.continuum}

In this appendix we review the two-dimensional Toda field theory in the continuum limit,
and set up some notations.

The conformal Toda field theory, or Toda field theory for short, is a two-dimensional CFT
associated with a simple Lie algebra $G$.\footnote{In Lie algebra notation this should be denoted by $G$. We here follow the literature of the Y-system and use the symbol $G$ for algebra.}. When we denote the rank of $G$ by $r$, the 
theory has $r$ scalar fields $\phi=(\phi^a)_{a=1, \ldots, r}$ parametrizing the Cartan subalgebra $H$ of $G$, and 
the Lagrangian is given by
\begin{align}
L= \int dx dt \, \sqrt{-g} \left[
\frac{1}{8\pi} \langle \partial^{\nu} \phi,   \partial_{\nu} \phi \rangle
+\mu  \sum_{a=1}^r  e^{b \langle \alpha_a,  \phi \rangle} 
+ \frac{1}{4\pi} \langle Q, \phi \rangle  R
\right]\;,
\label{eq.L}
\end{align}
In this expression, $\langle -, - \rangle$ is the
canonical pairing (Killing form) in $G$, with which we identify the 
elements of $G$ and its dual, and $\{\alpha_a\}_{a=1, \ldots, r}$ denotes a positive simple root.

In order for this theory to be conformal, 
the parameter $Q$ (background charge) should be related to another parameter $b$ as 
\begin{align}
Q:=\left( b  +\frac{1}{b} \right) \rho \;,
\end{align}
where $\rho$ is the Weyl vector:
\begin{align}
\rho=\frac{1}{2} \sum_{\alpha\in \Delta_{+}} \alpha \;,
\end{align}
with $\Delta_{+}$ the set of positive roots.
This model has a central charge 
\begin{align} 
c=r \left[ 1+h_G(h_G+1)\left(b+\frac{1}{b} \right)^2 \right] \;.
\label{c_Toda_G}
\end{align}
where $h_G$ is the (dual) Coxeter number of $G$.
For the case $G=A_{N-1}$, we have $h_G=N$ and this reduces to
\begin{align} 
c=(N-1) \left[ 1+N(N+1)\left(b+\frac{1}{b} \right)^2 \right] \;.
\label{c_Toda}
\end{align}


We can also choose $G$ to be an infinite-dimensional (untwisted or twisted) affine Lie algebra.
In this case, we have an extra affine simple root, $\alpha_0$, and 
the summation over $a$ in \eqref{eq.L} should now include $a=0$.
Physics in this case is very different, since the theory is 
a massive perturbation of a CFT and is non-conformal.


\section{Quantum Cluster Algebra}\label{sec.cluster}


For the convenience of the reader we here include minimal summary of quantum cluster algebras. The contents of this section is a simplified version of the appendices B
and C in \cite{Gang:2015wya}.\footnote{There is one difference in notation: $q$ here is $q^{1/2}$ in \cite{Gang:2015wya}. We have chosen this convention to remove the square roots from the time evolution rules \eqref{eq.evolve}.}

Let us begin with a quiver $Q$, {\it i.e.}, a finite oriented graph.
We denote its vertices by $i,j, \ldots \in I$.
Let us define an anti-symmetric matrix $\{ Q_{i,j}\}_{i,j\in I}$ as in \eqref{Q_def}.
The quivers discussed in this paper 
has no loops and oriented $2$-cycles,
and hence the quiver $Q$ can be identified with the anti-symmetric matrix
$\{ Q_{i,j}\}_{i,j\in I}$.

Given a vertex $k$, we define a new quiver $\mu_k Q$ (mutation of $Q$
at vertex $k$) by 
\begin{align}
(\mu_kQ)_{ij}:=
\begin{cases}
-Q_{ij} &  \text{($i=k$ or $j=k$)} \ , \\
Q_{ij} + [Q{}_{ik}]_+[Q_{kj}]_+ - [Q_{jk}]_+[Q_{ki}]_+ &
 \text{$(i,j\neq k$)} \ ,
\end{cases}
\label{Qmutate}
\end{align}
with $[x]_+:=\textrm{max}(x,0)$.

Given a quiver $Q=\{ Q_{i,j}\}_{i,j\in I}$, we associate quantum $y$-variable $\mathsf{y}_i$
for each vertex $i\in I$, and we impose the commutation relation \eqref{xCCR}.
The non-commutativity parameter is $q=e^{\hbar}$, with  ``Planck constant'' $\hbar$.
 
Mutation $\hat{\mu}_k$ acts on these quantum $y$-variables as
\begin{align}
\begin{split}
&\hat{\mu}_{k}\, \sfy_{i}\, \hat{\mu}_k^{-1} =q^{Q_{ik} [Q_{ik}]_{+}}\sfy_{i} \sfy_k^{[Q_{ik}]_{+}}
\prod_{m=1}^{|Q_{ki}|}\left(1+q^{\mathrm{sgn}(Q_{ki})(2m-1)}\sfy_{k}^{-1}\right)^{-\mathrm{sgn}(Q_{ki})}  \;.
\label{quantum mutation}
\end{split}
\end{align}
This can be represented as an operator
\begin{align}
\hat{\mu}_k =\psi_{\hbar}\left(\mathsf{Y}_k+i \pi b^2+i \pi \right) \hat{P}_k\;.
\label{mu_k}
\end{align}
Here $\psi_{\hbar}(x)$ is a quantum dilogarithm function 
satisfying the difference equation
\begin{align}
\begin{split}
&\psi_{\hbar}(z+ 2\pi i b^2)=(1-e^{-z})\, \psi_{\hbar}(z)  \;, 
\end{split}
\label{psi_difference_eq}
\end{align}
and the hermitian operator $\hat{P}_k$ give a transformation properties of (the logarithm of) the so-called tropical version of $y$-variables:
\begin{align}
\hat{P}_k (\mathsf{y}_i):=\hat{P}_k \mathsf{y}_i \hat{P}_k^{-1} = 
\left\{\begin{array}{cc} \mathsf{y}_k^{-1} & \quad\quad i=k \\ 
q^{Q_{ik} [Q_{ik}]_{+}} \mathsf{y}_i \mathsf{y}_k^{[Q_{ik}]_+}   & \quad  \quad  i\neq k \end{array}\right.  \;.
\end{align}
%

\parskip=-2pt

\bibliographystyle{nb}
\bibliography{chaos}

\end{document}